\documentclass[rnote]{aa}

\usepackage{graphicx}
\usepackage{natbib}
\bibpunct{(}{)}{;}{a}{}{,}

\graphicspath{{figs/}}

\usepackage{txfonts}

\newcommand\rg{\ensuremath{r_\mathrm{g}}} 
\newcommand\rx{\texttt{relxill}}

\begin{document}

   \title{Normalizing a relativistic model of X-ray reflection}

   \subtitle{ Definition of the reflection fraction and its
     implementation in \texttt{relxill} }

   \author{T. Dauser\inst{1}\thanks{thomas.dauser@sternwarte.uni-erlangen.de}
     \and J. Garc\'ia\inst{2} \and D.~J.~Walton\inst{3}$^{,}$\inst{4} \and
     W.~Eikmann\inst{1} \and T.~Kallman\inst{5} \and
     J.~McClintock\inst{2} \and J. Wilms\inst{1} }

   \institute{Remeis Observatory \& ECAP, Universit\"at
     Erlangen-N\"urnberg, Sternwartstr.~7, 96049 Bamberg, Germany \and Harvard-Smithsonian Center for
     Astrophysics, 60 Garden Street, Cambridge, MA 02138, USA \and
     Jet Propulsion Laboratory, California Institute of Technology, Pasadena, CA 91109, USA \and 
     Cahill Center for Astronomy and Astrophysics, California
     Institute of Technology, Pasadena, CA 91125, USA \and 
    X-ray Astrophysics Laboratory, NASA Goddard Space Flight Center, 
    Greenbelt, MD 20771, USA} 

   \date{Received ????; accepted ????}

   \abstract{}{ The only relativistic reflection model that implements
     a parameter relating the intensity incident on an accretion disk
     to the observed intensity is \rx. The parameter used in earlier
     versions of this model, referred to as the reflection strength,
     is unsatisfactory;  it has been superseded by a parameter that
     provides insight into the accretion geometry, namely the
     reflection fraction. The reflection fraction is defined as the
     ratio of the coronal intensity illuminating the disk to the coronal
     intensity that reaches the observer.

}{ The \rx\ model combines a general relativistic ray-tracing code and a
  photoionization code to compute the component of radiation reflected
  from an accretion that is illuminated by an external source. The
  reflection fraction is a particularly important parameter for
  relativistic models with well-defined geometry, such as the lamp post  
  model, which is a focus of this paper.

}{ Relativistic spectra are compared for three inclinations and for four
  values of the key parameter of the lamp post model, namely the height
  above the black hole of the illuminating, on-axis point source. In all
  cases, the strongest reflection is produced for low source heights and
  high spin. A low-spin black hole is shown to be incapable of producing
  enhanced relativistic reflection. Results for the \rx\ model are
  compared to those obtained with other models and a Monte Carlo
  simulation.

}{ Fitting data by using the \rx\ model and the recently implemented
     reflection fraction, the geometry of a system can be
     constrained. The reflection fraction is independent of system
     parameters such as inclination and black hole spin. The
     reflection-fraction parameter was implemented with the name
     \texttt{refl\_frac} in all flavours of the \rx\ model, and the
     non-relativistic reflection model \texttt{xillver}, in v0.4a (18
     January 2016).

}

\maketitle

\section{Introduction}

An important issue in the study of active galactic nuclei (AGNs) and
Galactic black hole binaries is that of the nature of the accretion
flow in the vicinity of the central accretion disk.  This flow is
usually modelled as a hot corona sandwiching the disk \citep[see,
  e.g.,][]{Haardt1991,Stern1995a}, but it may be a more complicated
structure, such as an outflow. The goal is to constrain the geometry
of this flow and its physical properties.  The chief means of addressing
this problem is by studying the spectrum of X-rays ``reflected'' from
the optically thick accretion disk due to its illumination by hard
X-rays produced in a surrounding corona \citep[see, e.g., the early
  work by][]{George1991a,Matt1991a}. The result is a rich spectrum of
radiative recombination continua, absorption edges, and fluorescent
lines, most notably the Fe K complex in the 6--8\,keV energy range
\citep{Matt1992a}. Determining the intensity of the reflected spectrum
relative to the spectrum that illuminates the disk can provide
important constraints on the geometry of the corona. In the simplest
case of Euclidean geometry, it is straightforward to parameterize the
relative intensity of the reflected component of emission. It is
proportional to the fraction of the disk that is covered by the
corona, which is the assumption underlying the widely used reflection
model \texttt{pexrav} \citep{Magdziarz1995a}.

For astrophysical black holes, however, the problem of parameterizing
the relative intensity of the reflected component is far more
complicated. The relative intensity and  the features in the
reflected spectrum are strongly affected by relativistic effects such
as light bending, special relativistic Doppler boosting, and
gravitational redshift. Relativistic reflection is commonly observed in
both AGNs \citep[see,
e.g.,][]{wilms2001a,Fabian2004a,Dauser2012a,Risaliti2013a,Walton2013a},
and in Galactic black holes such as Cyg~X-1 \citep[see,
e.g.,][]{Fabian1989,Duro2011a,Tomsick2014a,Parker2015a} and GX~339$-$4
\citep[see, e.g.,][]{Miller2008a,Garcia2015b}. In the presence of
relativistic effects, the relationship between the relative intensity of
the reflected spectrum and the geometry of the accretion flow becomes
complex and non-linear. For example, in the extreme limit where the
reflected component is dominant, the relative strength of this component
can be used to constrain the spin of a black hole
\citep[][]{Dauser2014a,Parker2014a}.

The purpose of this Research Note is to provide a clear definition of
a reflection-fraction parameter that captures the complex relationship
between the strength of the reflection signal and the geometry (which
has often been loosely referred to as the reflection fraction ; see,
e.g., \citealt{Walton2013a,Keck2015a}). We use the \rx\ model, which
is currently the only relativistic reflection model that implements a
reflection-fraction parameter. For a specific geometry, namely an
on-axis and isotropic point source, and by thorough consideration of
light-bending effects, we precisely define a normalization parameter,
the reflection fraction, which relates the incident and observed
spectra to the geometry of the system.

\section{Definitions}\label{sec:definition}

We define two principal quantities that serve to normalize the
observed spectrum relative to the coronal spectrum incident on the
disk: The reflection strength $R_\mathrm{s}$ and the reflection
fraction $R_\mathrm{f}$. Although they differ fundamentally, we show
that they are nevertheless related. Our focus is on $R_\mathrm{f}$ in
the relativistic case, and on its implementation in the \rx\ model.

\begin{figure*}
  \includegraphics[width=\textwidth]{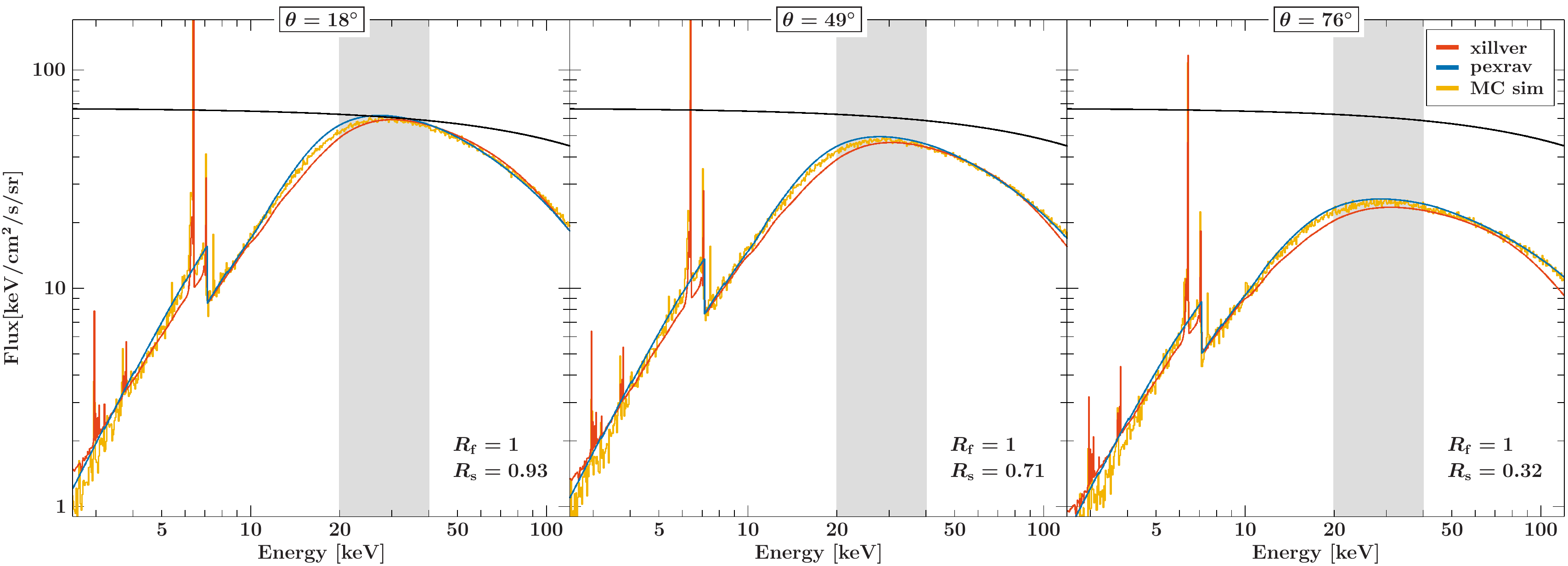}
  \caption{ Reflection spectra in the non-relativistic case and a neutral
    accretion disk for three models;  the inclination angle
    increases from left to right. The incident power-law spectrum is
    the black curve; the \texttt{xillver}, \texttt{pexrav}, and a
    Monte Carlo simulation are shown as red, blue, and yellow curves,
    respectively. We use \texttt{wilm} abundances \citep{Wilms2000a},
    except for \texttt{xillver} for which we use \texttt{grsa} abundances
    \citep{Grevesse1998a}. The highlighted area shows the energy band of
    the Compton hump, which is used to calculate the reflection
    strength. }
  \label{fig:norm_xillver}
\end{figure*}

\subsection{Reflection strength $R_\mathrm{s}$}
\label{sec:refl-fract-s_mathrmf}

The simple approach to parameterizing the strength of the reflection is
to use the ratio of the observed fluxes of the reflected component and
the incident component in some specified energy band. To this end, we
define the reflection strength to be this ratio in the
20--40\,keV band, a definition that has been used by e.g.,
\citet{Keck2015a} and \citet{Tao2015a}. This energy range, which
encompasses the peak of the Compton hump, is a good choice because the
reflection spectrum is dominated there by Compton scattering and
therefore depends weakly on the Fe abundance or ionization state of
the reflector. We note that some authors use a similar definition, but
employ a wider energy band \citep[see, e.g.,
][]{Wilkins2015b,Fuerst2015a}.

\subsection{Reflection fraction $R_\mathrm{f}$: The non-relativistic
Case}
\label{sec:refl-fract-r_mathrmf}

One disadvantage of the reflection strength, $R_\mathrm{s}$, is its
strong dependence on the inclination of the system, which makes it
difficult to relate this observable to the geometry of the
illuminating source. We therefore define a different quantity that is
independent of inclination and the condition of the reflector, namely
{\it the reflection fraction, which is the ratio of the coronal
  intensity that illuminates the disk to the coronal intensity that
  reaches the observer.} For a semi-infinite slab (i.e., a $2\,\pi$
accretion disk) and $R_\mathrm{f}=1$, the intensity of the coronal
component that illuminates the disk is the same as that seen by the
observer. In the non-relativistic case, this is the standard assumption
built into such widely used reflection models such as \texttt{pexrav}
\citep{Magdziarz1995a}, \texttt{reflionx} \citep{Ross2005a}, and
\texttt{xillver} \citep{Garcia2013a}. We note that \texttt{xillver}
and \texttt{pexrav} include a parameter that characterizes the
strength of the reflection spectrum, but that \texttt{reflionx} does
not.

Figure~\ref{fig:norm_xillver} shows reflection spectra computed using
the models \texttt{pexrav} \citep{Magdziarz1995a}, \texttt{xillver}
\citep{Garcia2013a}, and a Monte Carlo code for three inclination angles
$\theta$ (defined with respect to the normal of the accretion disk). All
the simulations are for the standard lamp post geometry: i.e., an
on-axis, isotropic point source, which is emitting the power-law
spectrum plotted in the figure. The spectrum illuminating the disk is
the same as that seen by the observer, i.e., $R_\mathrm{f}=1$.

In the limited bandpass considered in Figure~\ref{fig:norm_xillver}
(i.e., the X-ray band), the reflected flux decreases with increasing
inclination because flux is redistributed to energies below 100\,eV
\citep[e.g., see][]{Garcia2013a}. For the \texttt{xillver} model, the
reflection strength $R_\mathrm{s}$ for $\theta=18^\circ$, $49^\circ$,
and $76^\circ$ is 0.93, 0.71, and 0.32, respectively, with very similar
values for the other models. The modest differences between the models
are due to the use of different abundances and  to the approximation
used in \texttt{xillver}'s treatment of Compton scattering, which limits
its applicability to energies below approximately 100\,keV.

\subsection{Relativistic reflection}
\label{sec:rel_refl}

We now focus on the relevant and interesting case of relativistic
reflection with the spectrum blurred by gravitational redshift and by
Doppler and light-bending effects, a subject that has been widely
studied \citep[see, e.g.,][ and the review by
\citealp{Middleton2015a}]{Fabian1989,Laor1991,Dauser2010a}.  However,
the only relativistic model that parameterizes the relative strength of
the reflected spectrum is \rx\ \citep{Dauser2014a}. The model \rx\
combines our reflection code \texttt{xillver}
\citep{Garcia2010a,Garcia2013a} and the relativistic ray tracing code
\texttt{relline} \citep{Dauser2010a,Dauser2013a}.

The definition of the reflection-fraction parameter $R_\mathrm{f}$ in
the \rx\ model is identical to that given in
Sect.~\ref{sec:refl-fract-r_mathrmf} for the \texttt{xillver} and
\texttt{pexrav} models\footnote{ \citet{Basak2016a} propose an
  alternative definition, namely that the reflection fraction is the
  ratio of the coronal flux emitted towards the accretion disk to the
  coronal flux emitted towards the observer. The substantial
  difference to our definition is that only the direction of emission
  is used, not including the directional change due to light-bending;
  therefore, if the photon actually arrives at infinity or hits the
  accretion disk, it implies that $R_\mathrm{f}=1$ for any
  isotropically emitting lamp post source.} . There is a crucial
  complication that results from the effects of light bending: In
  order to precisely define $R_\mathrm{f}$ one must specify the
  geometry of the illuminating source because the observer no longer
  sees the same spectrum that illuminates the disk. For example, many
  photons initially directed toward infinity will strike the disk,
  thereby boosting the value of $R_\mathrm{f}$. Meanwhile, photons
  captured by the black hole or crossing the midplane beyond the outer
  radius of the disk are disregarded when computing
  $R_\mathrm{f}$. Because the black hole's gravity preferentially
  bends light rays back toward the disk and away from the observer,
  values of $R_\mathrm{f}>1$ are the norm, as illustrated by
  \citet[][their Fig.~2]{Dauser2014a}\footnote{While results obtained
    using the non-relativistic model pexrav cannot be directly
    compared to those obtained using \rx, the definition of
    $R_\mathrm{f}$ is the same for both models
    (Sect.~\ref{sec:refl-fract-r_mathrmf}).}.

\begin{figure*}
  \includegraphics[width=\textwidth]{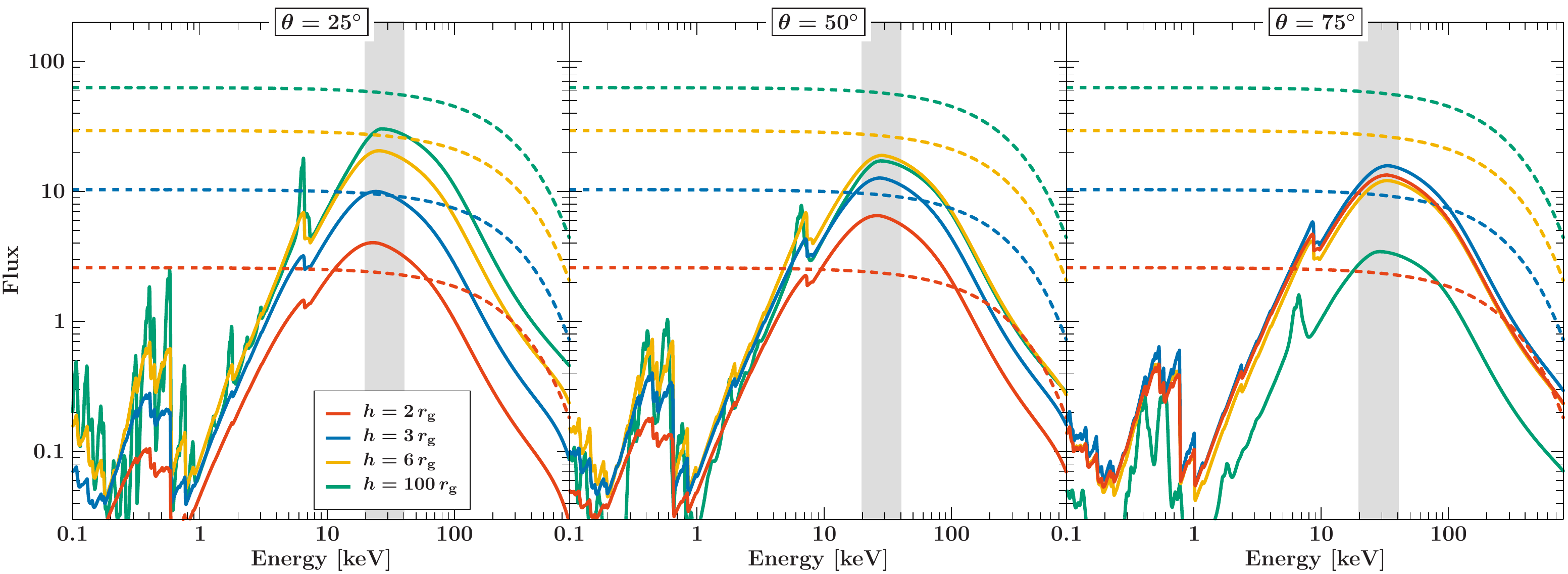}
  \caption{ Relativistic reflection spectra for accretion disks that are
    all illuminated by an isotropic on-axis point source with precisely
    the same luminosity in all cases.  Inclination increases to the 
    right;  each panel shows four models corresponding to four values
    of lamp post height $h$ ranging from 2~$r_{\rm g}$ to 100~$r_{\rm
      g}$. The dashed lines show the incident power-law spectra, while
    the solid lines depict the observed reflection spectra. For all
    models, the spin is $a=0.998$, the power-law index $\Gamma=2$, the
    high energy cutoff $E_\mathrm{cut}=300$\,keV, and the gas is
    neutral ($\log\xi=0$). The highlighted area shows the energy band of
    the Compton hump used in computing the reflection strength. }
  \label{fig:relat_mod}
\end{figure*}

Figure~\ref{fig:relat_mod} shows, for each of three values of
inclination, a set of reflected spectra and the corresponding spectra of
the power-law that illuminates the disk. The emissivity profiles are
those appropriate for a lamp post geometry \citep[see,
e.g.,][]{Martocchia1996a,Martocchia2002a,Dauser2013a}, where an on-axis
source is located above the black hole at heights of $h=2$, 3\rg, 6\rg, and
$100\rg$ ($r_{\rm g} = GM/c^2$), which correspond to reflection
fractions of $R_\mathrm{f}=5.9$ of 3.3, 1.8, and 0.8, respectively
\citep[see][for more details and other parameter
combinations]{Dauser2014a}. The luminosity of the point source is the
same in all cases. Therefore, the flux in the incident power-law spectra
(dashed lines) decreases with $h$ as the gravitational redshift
increases.

As the figure shows, the reflection spectrum depends strongly on
inclination. Interestingly, for small values of $h$ the reflection
strength $R_\mathrm{s}$ increases with inclination, while for large
$h$ it decreases. One reason for this effect is that redshift and
Doppler-boost effects depend strongly on the radial velocity of the
disk material, and hence on the inclination of the disk.  A second
reason is that the emission angle, the angle at which the observer
views the disk, is altered by relativistic effects that diminish with
radius \citep[see][]{Garcia2014a}. The emission angle strongly affects
the reflected spectrum because the spectrum is dominated by those
portions of the disk that are viewed face-on, i.e., the regions that are near
the black hole that exhibit the strongest Doppler boosting. Also
important is the height $h$ of the point source above the black hole:
The relativistic effects are strong for a point source that is near,
but for a source located far from the hole the relativistic effects
are muted and the dependence of the reflected spectrum on inclination
becomes similar to that for the non-relativistic case
(Sect.~\ref{sec:refl-fract-r_mathrmf}).

In summary, for relativistic reflection in the lamp post geometry (or
any other fully specified geometry), we can compute the reflection
fraction. As the lamp post geometry is the only geometry fulfilling
these conditions, and this geometry is implemented in relativistic
reflection models, it is the focus of the following discussion.  By
comparing the observed value of the reflection fraction to the model
values, we are able to place constraints on the geometry of the
system, specifically the height $h$ of the illuminating source and the
disk inclination angle. We close this section with two
conclusions. First, for lamp post geometry and an isotropic point
source of constant luminosity, the strongest reflection is produced
for small values of $h$ and high inclination where relativistic
effects are strong, and for large values of $h$ and low inclination
where they are weak. Second, for a constant value of the reflection
fraction $R_\mathrm{f}$, the ratio of the incident to the reflected
flux (i.e., the reflection strength $R_\mathrm{s}$) is strongly
dependent on $h$ and inclination.

\section{ Relationship of $R_\mathrm{f}$ to $R_\mathrm{s}$}
\label{sec:relationship}

\begin{figure*}
  \centering
  \includegraphics[width=\textwidth]{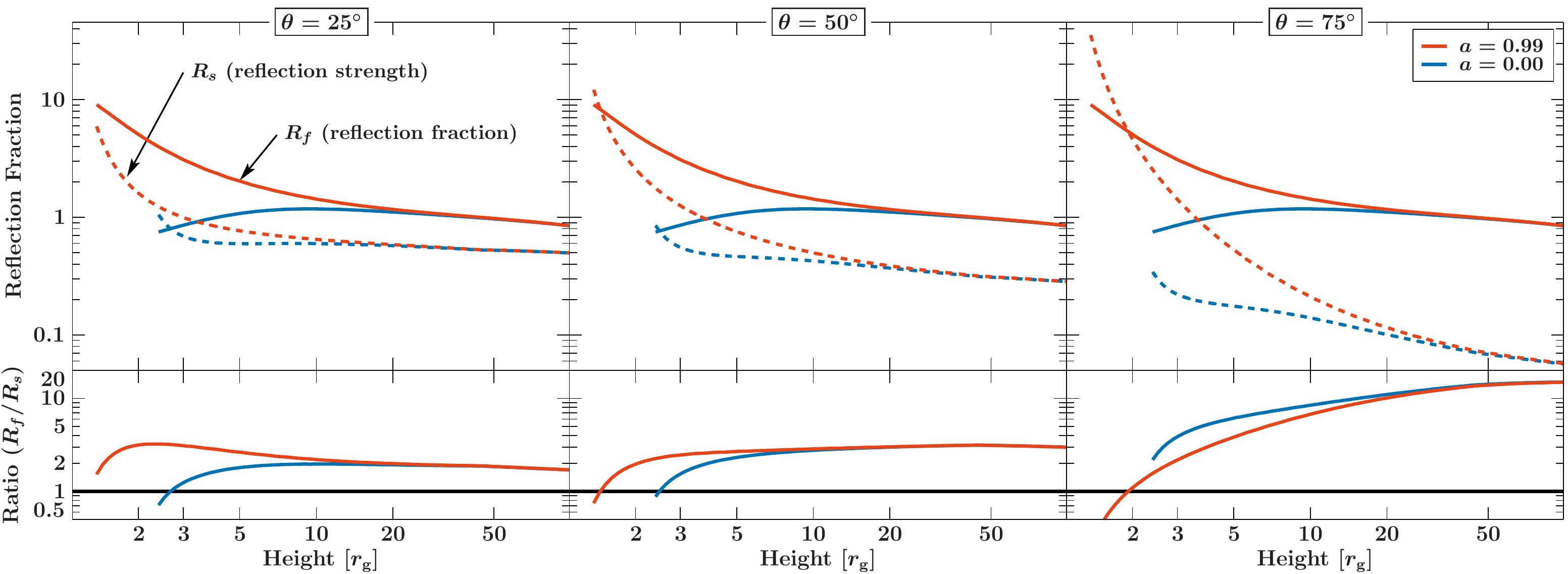}
  \caption{Comparison of the reflection fraction $R_\mathrm{f}$ (solid
    lines) and the reflection strength $R_\mathrm{s}$ (dashed lines) for
    increasing lamp post geometry (from left to right) and
    two values of spin, $a=0$ (blue curves) and $a=0.99$ (red
    curves). The ratio of $R_\mathrm{f}$ to $R_\mathrm{s}$ is shown in the
    lower panels.}
  \label{fig:ratio}
\end{figure*}

We now consider the relationship of the reflection fraction
$R_\mathrm{f}$ to the reflection strength $R_\mathrm{s}$. This latter
parameter is a straightforward observable and its value is meaningful,
and sometimes quoted, when fitting data using non-relativistic models
such as \texttt{pexrav} and \texttt{pexrav}
(Sect.~\ref{sec:refl-fract-s_mathrmf}). However, its value is very
limited in modelling strongly relativistic systems, e.g., when using \rx.

Figure~\ref{fig:ratio} illustrates the relationship between
$R_\mathrm{f}$ and $R_\mathrm{s}$ for three values of inclination as a
function of the height $h$ of the lamp post. The quantities themselves
are plotted in the top panels, and their ratio is plotted in the lower
panels.  These results are based on the same ray-tracing calculations
used to generate the spectra plotted in Fig.~\ref{fig:relat_mod}. For
most cases, $R_\mathrm{f}>R_\mathrm{s}$. For low and moderate
inclinations, this ratio is roughly a factor of two, and decreases rather
abruptly for small values of $h$. The high inclination case is quite
different: For large $h$, $R_\mathrm{f}$ is about an order of magnitude
greater than $R_\mathrm{s}$, while for small $h$ and high spin the
ratio plummets and $R_\mathrm{s}$ greatly exceeds $R_\mathrm{f}$.

It is interesting to compare the different results for low-spin and
high-spin black holes. In the low-spin case, both $R_\mathrm{f}$ and
$R_\mathrm{s}$ are $\lesssim1$. For rapidly spinning black holes and
$h\lesssim5$, on the other hand, both reflection parameters are much
larger, making their corresponding reflection spectra relatively more
prominent. This is a plausible observational selection effect that
helps to explain the observed prevalence of high-spin black holes
\citep[see, e.g.,][]{Reynolds2014a,Middleton2015a}, but we also note
that this effect can be partly explained by a larger radiative
efficiency of the rapidly rotating black holes
\citep[see][]{Vasudevan2016a}.

Because \rx\ is currently the only relativistic model that implements a
reflection-fraction parameter, we have no point of comparison. We note,
however, that values of reflection fraction have been quoted for the
relativistic model \texttt{reflionx} \citep[see, e.g.,][]{Keck2015a},
even though this parameter is not explicitly included in the model. In
these cases, the values quoted are for the reflection strength
$R_\mathrm{s}$ (and not $R_\mathrm{f}$).

\section{ Reflection fraction in the \texttt{relxill} model}

In the following we present the implementation of $R_\mathrm{f}$ (in
place of $R_\mathrm{s}$) in the example \texttt{relxill} model and
also state our reasons for this choice. First, it is applied to the
lamp post geometry, the only geometry implemented in a relativistic
reflection model, which is fully specified. Second, we emphasize the
uncertain interpretation of $R_\mathrm{f}$ for the standard power-law
emissivity version of \texttt{relxill}, presenting a case which does
not provide a strict geometrical definition. For completness, the
specific normalizations of the \texttt{xillver} and \texttt{relxill}
models are spelled out in Appendix~\ref{sect:appendix}.

\subsection{ Adoption of the reflection fraction }
\label{sec:refl-fract-textttr}

We have consistently used the parameter $R_\mathrm{f}$ to quantify the
normalization of the reflected component in all flavours of the \rx\
model since we released version 0.4a on 2016 January 18. Previously, we
used the reflection strength, which is not simply related to
$R_\mathrm{f}$ (Sect.~\ref{sec:relationship}). Figure~\ref{fig:ratio}
provides a rough idea of the relationship between the two parameters.

The principal reason for adopting $R_\mathrm{f}$ is that for the lamp post
version of \rx\ this observable allows   constraints to be placed on the
geometry of the system, specifically the lamp post height $h$ and
inclination angle (Sect.~\ref{sec:rel_refl}; \citealt{Dauser2014a}). The
reflection strength, by comparison, does not provide insight into the
geometry of the system. Furthermore, the reflection strength depends on
such parameters as the inclination and the black hole's spin, while
$R_\mathrm{f}$ is independent of these parameters.

Moreover, as demonstrated by \citet{Dauser2014a}, when fitting
observational data with the lamp post version of \rx, 
additional constraints can be obtained on the spin parameter by excluding values of
$R_\mathrm{f}$ that are unrealistic. This is possible because
$R_\mathrm{f}$ is closely tied to the accretion geometry, while at the
same time it is computed for the observed disk spectrum (i.e.,
ignoring relativistic effects on light rays traveling from
the disk to the observer). As an example of a constraint enabled by
$R_\mathrm{f}$, \citet{Dauser2014a} show that for the larger inner-disk
radius of a low-spin black hole, large values of $R_\mathrm{f}$ are
excluded because a substantial fraction of the photons are captured by
the black hole.

The parameter $R_\mathrm{f}$ was implemented for the \texttt{xillver}
model at the same time as for the \rx\ model. Conveniently, the
widely used \texttt{pexrav} model employs the same normalization, which
allows a direct comparison between the two non-relativistic reflection
models \texttt{pexrav} and \texttt{xillver}.

\subsection{{Standard \texttt{relxill} model with emissivity index}}

A principal virtue of the lamp post version of \rx\ is that its geometry
is completely defined. This is not true of the standard version of \rx,
which follows the venerable tradition of describing the illumination
profile of the disk by a broken power law \citep{Fabian1989}. In this
case, the geometry of the illuminating source is undefined because many
conceivable geometries could produce the same power-law illumination
profile. In defining $R_\mathrm{f}$ for the standard model, we assume
that the geometry of the illuminating source   is a razor-thin
layer that hugs the entire disk. Hence, unlike the lamp post geometry,
the photons illuminating the disk are not shifted in energy, and
light-bending is irrelevant.

This simplistic geometry is very unlikely to represent physical systems,
and it is particularly inappropriate for models with steep emissivity
profiles in the inner-disk region. Given that a unique geometry cannot
be specified for the emissivity-index model, and given the simplistic
geometry we have adopted, fitted values of $R_\mathrm{f}$ for this
version of the \rx\ model are of quite limited value when assessing the
geometry of a system.

\section{Summary and conclusions} 

We have discussed two normalization parameters for use in models of
X-ray reflection. The first of these, the reflection strength
$R_\mathrm{s}$,  is the ratio of the flux incident on the disk to
the reflected component of flux in the 20--40~keV band. One disadvantage of the reflection strength is its dependence on system
parameters such as inclination and black hole spin. Furthermore, it does
not provide insight into the geometry of the system.

Because of these flaws, we adopted a new normalization parameter, the
reflection fraction $R_\mathrm{f}$, which was first implemented in
both the \rx\ and \texttt{xillver} models in v0.4a (18 January 2016)
use the same date format throughout the paper ; the parameter name in
the models is \texttt{refl\_frac}. The reflection fraction is defined
as the ratio of the coronal intensity that illuminates the disk to the
coronal intensity that reaches the observer. In computing
$R_\mathrm{f}$, all relativistic effects are included for light rays
traveling from the illuminating source to the disk, but these effects
do not act on photons traveling from the disk to the observer.  A
principal virtue of $R_\mathrm{f}$ is that if the geometry is
specified, then the geometrical parameters can be constrained by
observation, as in the case of the lamp post scenario. Another virtue
is that $R_\mathrm{f}$ does not depend on the system parameters of
inclination and black hole spin.

\begin{acknowledgements}
  We thank John E.~Davis for the development of the \textsc{SLxfig}
  module used to prepare the figures in this research note and Andrzej
  Zdziarski for helpful comments. This research has made use of ISIS
  functions (ISISscripts) provided by ECAP/Remeis observatory and MIT
  (http://www.sternwarte.uni-erlangen.de/isis/). We also thank the
  anonymous referee for comments that improved the manuscript.
\end{acknowledgements}

\bibliographystyle{jwaabib}     
\bibliography{mnemonic,aa_abbrv,local} 

\begin{thebibliography}{}

\bibitem[\protect\astroncite{{Ballantyne}}{2004}]{Ballantyne2004a}
{Ballantyne} D.R.,  2004, MNRAS 351, 57

\bibitem[\protect\astroncite{{Ballantyne} et~al.}{2001}]{Ballantyne2001a}
{Ballantyne} D.R., {Iwasawa} K., {Fabian} A.C.,  2001, MNRAS 323, 506

\bibitem[\protect\astroncite{{Basak} \& {Zdziarski}}{2015}]{Basak2016a}
{Basak} R., {Zdziarski} A.A.,  2015, MNRAS (submitted, arXiv:1512.01833)

\bibitem[\protect\astroncite{{Dauser} et~al.}{2014}]{Dauser2014a}
{Dauser} T., {Garc{\'{\i}}a} J., {Parker} M., et~al., 2014, MNRAS 444, L100

\bibitem[\protect\astroncite{{Dauser} et~al.}{2013}]{Dauser2013a}
{Dauser} T., {Garc{\'{\i}}a} J., {Wilms} J., et~al., 2013, MNRAS  687

\bibitem[\protect\astroncite{{Dauser} et~al.}{2012}]{Dauser2012a}
{Dauser} T., {Svoboda} J., {Schartel} N., et~al., 2012, MNRAS 422, 1914

\bibitem[\protect\astroncite{{Dauser} et~al.}{2010}]{Dauser2010a}
{Dauser} T., {Wilms} J., {Reynolds} C.S., {Brenneman} L.W.,  2010, MNRAS 409,
  1534

\bibitem[\protect\astroncite{{Duro} et~al.}{2011}]{Duro2011a}
{Duro} R., {Dauser} T., {Wilms} J., et~al., 2011, A\&A 533, L3

\bibitem[\protect\astroncite{{Fabian} et~al.}{2004}]{Fabian2004a}
{Fabian} A.C., {Miniutti} G., {Gallo} L., et~al., 2004, MNRAS 353, 1071

\bibitem[\protect\astroncite{{Fabian} et~al.}{1989}]{Fabian1989}
{Fabian} A.C., {Rees} M.J., {Stella} L., {White} N.E.,  1989, MNRAS 238, 729

\bibitem[\protect\astroncite{{F{\"u}rst} et~al.}{2015}]{Fuerst2015a}
{F{\"u}rst} F., {Nowak} M.A., {Tomsick} J.A., et~al., 2015, ApJ 808, 122

\bibitem[\protect\astroncite{{Garc{\'{\i}}a} et~al.}{2014}]{Garcia2014a}
{Garc{\'{\i}}a} J., {Dauser} T., {Lohfink} A., et~al., 2014, ApJ 782, 76

\bibitem[\protect\astroncite{{Garc{\'{\i}}a} et~al.}{2013}]{Garcia2013a}
{Garc{\'{\i}}a} J., {Dauser} T., {Reynolds} C.S., et~al., 2013, ApJ 768, 146

\bibitem[\protect\astroncite{{Garc{\'{\i}}a} \& {Kallman}}{2010}]{Garcia2010a}
{Garc{\'{\i}}a} J., {Kallman} T.R.,  2010, ApJ 718, 695

\bibitem[\protect\astroncite{{Garc{\'{\i}}a} et~al.}{2015}]{Garcia2015b}
{Garc{\'{\i}}a} J.A., {Steiner} J.F., {McClintock} J.E., et~al., 2015, ApJ 813,
  84

\bibitem[\protect\astroncite{{George} \& {Fabian}}{1991}]{George1991a}
{George} I.M., {Fabian} A.C.,  1991, MNRAS 249, 352

\bibitem[\protect\astroncite{{Grevesse} \& {Sauval}}{1998}]{Grevesse1998a}
{Grevesse} N., {Sauval} A.J.,  1998, Space Sci. Rev. 85, 161

\bibitem[\protect\astroncite{{Haardt} \& {Maraschi}}{1991}]{Haardt1991}
{Haardt} F., {Maraschi} L.,  1991, ApJ 380, L51

\bibitem[\protect\astroncite{{Keck} et~al.}{2015}]{Keck2015a}
{Keck} M.L., {Brenneman} L.W., {Ballantyne} D.R., et~al., 2015, ApJ 806, 149

\bibitem[\protect\astroncite{{Laor}}{1991}]{Laor1991}
{Laor} A.,  1991, ApJ 376, 90

\bibitem[\protect\astroncite{{Magdziarz} \& {Zdziarski}}{1995}]{Magdziarz1995a}
{Magdziarz} P., {Zdziarski} A.A.,  1995, MNRAS 273, 837

\bibitem[\protect\astroncite{{Martocchia} \& {Matt}}{1996}]{Martocchia1996a}
{Martocchia} A., {Matt} G.,  1996, MNRAS 282, L53

\bibitem[\protect\astroncite{{Martocchia} et~al.}{2002}]{Martocchia2002a}
{Martocchia} A., {Matt} G., {Karas} V., et~al., 2002, A\&A 387, 215

\bibitem[\protect\astroncite{{Matt} et~al.}{1991}]{Matt1991a}
{Matt} G., {Perola} G.C., {Piro} L.,  1991, A\&A 247, 25

\bibitem[\protect\astroncite{{Matt} et~al.}{1992}]{Matt1992a}
{Matt} G., {Perola} G.C., {Piro} L., {Stella} L.,  1992, A\&A 257, 63

\bibitem[\protect\astroncite{{Middleton}}{2015}]{Middleton2015a}
{Middleton} M.,  2015, To be published in: "Astrophysics of Black Holes - From
  fundamental aspects to latest developments", Ed. Cosimo Bambi, Springer:
  Astrophysics and Space Science Library (arXiv:1507.06153)

\bibitem[\protect\astroncite{Miller et~al.}{2008}]{Miller2008a}
Miller J.M., Reynolds C.S., Fabian A.C., et~al., 2008, ApJ 679, L113

\bibitem[\protect\astroncite{{Parker} et~al.}{2015}]{Parker2015a}
{Parker} M.L., {Tomsick} J.A., {Miller} J.M., et~al., 2015, ApJ 808, 9

\bibitem[\protect\astroncite{{Parker} et~al.}{2014}]{Parker2014a}
{Parker} M.L., {Wilkins} D.R., {Fabian} A.C., et~al., 2014, MNRAS 443, 1723

\bibitem[\protect\astroncite{{Reynolds}}{2014}]{Reynolds2014a}
{Reynolds} C.S.,  2014, Space Sci. Rev. 183, 277

\bibitem[\protect\astroncite{{Risaliti} et~al.}{2013}]{Risaliti2013a}
{Risaliti} G., {Harrison} F.A., {Madsen} K.K., et~al., 2013, Nat 494, 449

\bibitem[\protect\astroncite{{Ross} \& {Fabian}}{2005}]{Ross2005a}
{Ross} R.R., {Fabian} A.C.,  2005, MNRAS 358, 211

\bibitem[\protect\astroncite{{Stern} et~al.}{1995}]{Stern1995a}
{Stern} B.E., {Poutanen} J., {Svensson} R., et~al., 1995, Astrophys. J., Lett.
  449, L13

\bibitem[\protect\astroncite{{Tao} et~al.}{2015}]{Tao2015a}
{Tao} L., {Tomsick} J.A., {Walton} D.J., et~al., 2015, ApJ 811, 51

\bibitem[\protect\astroncite{{Tomsick} et~al.}{2014}]{Tomsick2014a}
{Tomsick} J.A., {Nowak} M.A., {Parker} M., et~al., 2014, ApJ 780, 78

\bibitem[\protect\astroncite{{Vasudevan} et~al.}{2016}]{Vasudevan2016a}
{Vasudevan} R.V., {Fabian} A.C., {Reynolds} C.S., et~al., 2016, MNRAS

\bibitem[\protect\astroncite{{Walton} et~al.}{2013}]{Walton2013a}
{Walton} D.J., {Nardini} E., {Fabian} A.C., et~al., 2013, MNRAS 428, 2901

\bibitem[\protect\astroncite{{Wilkins} et~al.}{2015}]{Wilkins2015b}
{Wilkins} D.R., {Gallo} L.C., {Grupe} D., et~al., 2015, MNRAS 454, 4440

\bibitem[\protect\astroncite{{Wilms} et~al.}{2000}]{Wilms2000a}
{Wilms} J., {Allen} A., {McCray} R.,  2000, ApJ 542, 914

\bibitem[\protect\astroncite{Wilms et~al.}{2001}]{wilms2001a}
Wilms J., Reynolds C.S., Begelman M.C., et~al., 2001, MNRAS 328, L27

\end{thebibliography}

\appendix
\section{Normalization of the \texttt{xillver} and \texttt{relxill}
  models} 
\label{sect:appendix}

The \texttt{xillver} model is normalized for an incident spectrum
(currently, a cutoff power-law) with flux $F_X(E)$ such that

\begin{equation}
  \int_{0.1\mathrm{keV}}^{\mathrm{1MeV}} F_X(E)\, \mathrm{d}E = 10^{20} \frac{n
    \xi}{4 \pi} \quad,
\end{equation}

\noindent where the density and ionization parameter are fixed to the
values $n=10^{15}\,\mathrm{cm}^3$ and $\xi =
1~\mathrm{erg}\,\mathrm{cm}\,\mathrm{s}^{-1}$, respectively \citep[see
  also][]{Garcia2013a}. While the normalization of \rx\ (which is
based on \texttt{xillver}) is identical, the flux reaching the
observer differs because of the relativistic effects described above.

\end{document}